\documentclass[aps,prl,amsfonts,twocolumn,superscriptaddress,showpacs]{revtex4}
\usepackage{amssymb}
\usepackage{graphicx,hyperref,pict2e}
\usepackage{epsfig}

\begin{document}

\title{Interaction Effects on Wannier Functions of a Bose-Einstein Condensate  in  an Optical
Lattice and Implications for Bose-Hubbard Model}

\author{Z. X. Liang}
\affiliation{Department of physics, Centre for Nonlinear Studies,
and The Beijing-Hong Kong-Singapore Joint Centre for Nonlinear and
Complex (Hong Kong), Hong Kong Baptist University, Kowloon Tong,
Hong Kong, China}

\affiliation{Shenyang National Laboratory for Materials Science,
Institute of Metal Research, Chinese Academy of Sciences, Wenhua
Road 72, Shenyang 110016, China}

\author{BamBi Hu}
\affiliation{Department of physics, Centre for Nonlinear
Studies, and The Beijing-Hong Kong-Singapore Joint Centre for
Nonlinear and Complex (Hong Kong), Hong Kong Baptist University,
Kowloon Tong, Hong Kong, China}
\affiliation{Department of Physics, University of Houston,
Houston, TX 77204-5005, USA}

\author{Biao Wu}
\thanks{bwu@aphy.iphy.ac.cn}
\affiliation{Institute of Physics, Chinese Academy of Sciences,
Beijing 100190, China}
\date{\today}
\begin{abstract}
We show that one can properly take into account of the interaction
effects and construct a set of orthonormal Wannier functions for a
Bose-Einstein condensate in an optical lattice. These
interaction-dependent Wannier functions are used to compute  the
tunneling rate $J$ and the on-site repulsion $U$ in the Bose-Hubbard
model. Both parameters are found to be substantially different from
ones calculated with the single-particle Wannier functions.
Our numerical results of $U$ are found in good agreement
with the measured  on-site energy  in a recent
experiment [Campbell {\it et al.} Science {\bf 314}, 281 (2006)].
\end{abstract}
\pacs{03.75.Lm,05.30.Jp,71.15.-m,73.43.Nq}

\maketitle

The system of a Bose-Einstein condensate (BEC) in an optical lattice
has been one of the most exciting and studied systems in recent
years \cite{Blochrev1,Morsch,Blochrev2,Yukalov}. The system has
enabled experimentalists to observe for the first time many
interesting phenomena predicted a long time ago in condensed matter
physics. The most spectacular example is the observation of the
quantum phase transition from a superfluid to a Mott insulator in
such a BEC system \cite{Greiner,Xuk,Stoferle,Spielman1,Spielman2}.

To understand this periodic BEC system, one often uses the Wannier
functions and reduces the system to the famed Bose-Hubbard model
(BHM) \cite{Fisher,Jaksch,Oosten,Zwerger}. The  Hamiltonian of the
BHM is given by
\begin{equation}\label{BHM}
\hat{H}=-J\sum_{<i,j>}\hat{a}^{\dag}_i\hat{a}_j+
\frac{U}{2}\sum_i\hat{n}_i(\hat{n}_i-1),
\end{equation}
where $\hat{a}_i$ and $\hat{a}^{\dagger}_i$ are respectively the
bosonic annihilation and creation operators at the $i$th lattice
site and $\hat{n}_i=\hat{a}^{\dag}_i\hat{a}_i$.  The angle brackets
above indicate the summation over all nearest neighboring pairs. The
tunneling rate $J$ and the on-site repulsion $U$ can be computed
 with the Wannier function $w({\bf
r})$ \cite{Jaksch,Oosten,Zwerger}. For atoms of mass $m$ and
$s$-wave scattering length $a_s$, they are given by \cite{Jaksch}
\begin{equation}\label{J}
J=-\int d{\bf r} w^{*}({\bf r}-{\bf
r}_i)\left[-\frac{\hbar^2\nabla^2}{2m}+V_{latt}\left({\bf
r}\right)\right]w({\bf r}-{\bf r}_j),
\end{equation}
and
\begin{equation}\label{U}
U=\frac{4\pi \hbar^2 a_s}{m}\int d {\bf r}|w({\bf r})|^4,
\end{equation}
where ${\bf r}_i$ and ${\bf r}_j$ are the coordinates of a pair of
nearest neighboring sites.  The potential for the optical lattice
has the form $V_{latt}=V_0 E_R\sum_{j=1}^{D}\sin^2\left(q_B
x_j\right)$ with $q_B=\pi/d$ being the laser wave vector, $d$ the
lattice period and $V_0$ the laser intensity in units of the recoild
energy $E_R=\hbar^2q_B^2/2m$. $D=1,2,3$ is the
dimensionality of the lattice. As is well known \cite{Zwerger}, when
the filling factor, namely the average number $\langle
\hat{n}_i\rangle$ of bosons at one site,  is fixed, the physics of
the BHM is completely determined by its tunneling rate $J$ and
on-site repulsion $U$. Hence it is crucial to have accurate values
of $J$ and $U$ for a good description of the BEC system with the BHM.

So far, these two parameters are usually computed with the
single-particle Wannier function \cite{Jaksch,Oosten,Zwerger}. Such
treatment is good only in the low-filling regime \cite{Blochrev2}.
For higher fillings, due to stronger inter-atomic interactions, the
shape of the Wannier function is expected to distinguish
significantly from that of the single-particle Wannier function. As
a result, $J$ and $U$ are interaction-dependent. In particular, $J$
is more sensitive since it depends on the tails of the Wannier
function as seen in Eq. (\ref{J}).  This view is echoed in
literature. For example, Bloch {\it et al.} pointed out
\cite{Blochrev2}, {\it``For intermediate fillings, the Wannier
functions entering both the effective hopping matrix element $J$ and
on-site repulsion $U$ have to be adjusted to account for the
mean-field interaction"}. In a recent experiment by Campbell {\it et
al.} \cite{Campbell}, the one-site energy with the filling factors
increasing from one to five was observed to have a $27\%$ decrease.

To our best knowledge, there has been only one
systematic theoretical attack on this important problem. This was carried out by
Li {\it et al.} in Ref. \cite{Li}, where the authors constructed a
set of orthonormal interaction-dependent Wannier functions  with
Kohn's variational approach \cite{Kohn2}. This variational method
has one intrinsic shortcoming: the chosen trial
Wannier function may be quite different from the true Wannier function.
In Ref. \cite{Li},  Li {\it et al.} used
Gaussian functions as the trial functions. However, as pointed out
in Ref. \cite{Blochrev2,Kramers}, the Gaussian approximation is
 not good  for calculating the tunneling parameter
$J$.  This means that even the best Gaussian type Wannier functions
obtained by variation in Ref. \cite{Li} are not good enough.

In this Letter we show that a set of  Wannier functions can be
constructed from the Bloch states of the Gross-Pitaevskii equation
(GPE) for the periodic BEC system. These Wannier functions are
proved to be orthonormal and are therefore suitable for the use of
reduing the BEC system to the BHM. Moreover, these Wannier functions
are interaction-dependent by construction; we call them
nonlinear Wannier functions to distinguish from the single-particle
Wannier functions. That the
interaction effects are properly taken into account in these
nonlinear Wannier functions roots in the  fact that the Bloch states
used to construct them minimizes the system energy under the
mean-field approximation. With these nonlinear Wannier functions,
we find that both the tunneling rate $J$ and the on-site
repulsion $U$ are substantially
affected by the mean-field interaction. For simplicity, the
construction and properites of the nonlinear Wannier functions are
illustrated in detail for one-dimensional optical lattice while the
results for the BHM parameters $J$ and $U$ are presented for all
dimensionality.

To define the nonlinear Wannier functions, we first introduce the
mean-field Bloch states $\psi_{\bf k}({\bf r})$
\cite{Wu,Diakonov,Bronski}, which satisfy the following
time-independent GPE \cite{Dalfovo}
\begin{equation}\label{GPE}
-\frac{\hbar^2\nabla^2}{2m}\psi_{{\bf k}}+V_{latt}\psi_{{\bf
k}}+gn_0|\psi_{{\bf k}}|^2\psi_{{\bf k}}=\mu({\bf
k})\psi_{{\bf k}}\,,
\end{equation}
where $g=4\pi\hbar^2 a_s/m$ and $n_0$ is the average BEC density.
The band index is omitted as we focus on the lowest Bloch
band. The nonlinear Wanniner functions are constructed from these
Bloch states as follows
\begin{equation}\label{Wannier}
w\left({\bf r}-{\bf
r}_i\right)=\frac{1}{|\Omega|}\int_{\Omega}\psi_{{\bf k}}({\bf
r})e^{-i{\bf k}\cdot {\bf r}_i} d{\bf k}\,,
\end{equation}
where the integration is over the first Brillouin zone and
$|\Omega|$ is its volume.  This is exactly the same way how the
single-particle Wannier function is constructed from Bloch states
\cite{Kohn,Ashcroft}. Because of the nonlinear term in Eq.
(\ref{GPE}), one may doubt whether the Wannier functions constructed
in such a way are orthonormal to each other and thus have any use.
This doubt can be cast aside immediately by observing that the Bloch
states defined in Eq. (\ref{GPE}) are orthonormal to each other,
i.e., $\int\psi^{*}_{{\bf k}^{'}}({\bf r})\psi_{\bf k}({\bf r})d{\bf
r}=\delta_{{\bf k}^{'}{\bf k}}$, despite the nonlinear term in Eq.
(\ref{GPE}).  With this, one can prove the orthonormality of the
nonlinear Wannier functions
\begin{equation}
\int w^{*}({\bf r}-{\bf r}_i)w({\bf r}-{\bf r}_j)d{\bf
r}=\delta_{ij}\,,
\end{equation}
where the integration is over the entire space.  There was a concern
in Ref. \cite{Li} that the definition in Eq. (\ref{Wannier}) would
fail because of the existence of a loop structure in the mean-field
Bloch band $\mu({\bf k})$ \cite{Wu,Diakonov,Seaman}. This concern is
not justified because the BHM is a single-band approximation and it
is a good description of the BEC system only when the band gap of
the lattice potential is much larger than the interatomic
interaction \cite{Jaksch}. The loop structure in the Bloch band
appears only for shallow optical lattices, where the BHM does not
apply.

We now use the 1D case to illustrate some key points in the
numerical computation of the nonlinear Wannier functions. The
complete  set of the Bloch functions $\psi_{k}(x)$ in the lowest
Bloch band can be obtained by solving Eq. (\ref{GPE}) with the same
method in Ref. \cite{Liang}. Since an arbitrary phase can be added
to each $\psi_{k}(x)$, to obtain a proper Wannier function via
Eq.(\ref{Wannier}), the numerical method must be designed to make
sure that the resulted $\psi_{k}(x)$ is analytic in $k$. Usually,
one also wants the Wannier function to be real and symmetric (or
antisymmetric). To achieve this, one should pay attention to  the
values of $\psi_{0}(0)$ and $\psi_{\pi/d}(0)$ \cite{Kohn2}.  ({\it
i}) If both $\psi_{0}(0)$ and $\psi_{\pi/d}(0)$ are nonzero, the
phase of the Bloch function must be chosen such that $\psi_{k}(0)$
is real. ({\it ii}) If both $\psi_{0}(0)$ and $\psi_{\pi/d}(0)$
vanish, the phase must be chosen such that $\psi_{k}(0)$ is purely
imaginary. ({\it iii}) If only one of $\psi_{0}(0)$ and
$\psi_{\pi/d}(0)$ is zero, one can shift the origin in the $x$ space
by half of the lattice constant. With the new origin, one is then
back to either case ({\it i}) or ({\it ii}). One can prove that
 (1) if $\psi_{0}(0)\neq0$, $w(-x)=w(x)$ and $w^{*}(x)=w(x)$, that is,
 the Wannier function $w(x)$ is symmetric
about $x=0$ and real; (2) if $\psi_{0}(0)=0$, $w(x)$ is
antisymmetric about $x=0$ and real. Following Kohn's strategy
\cite{Kohn2}, one can also prove that no other choices of phases in
$\psi_{k}(x)$ can  lead to nonlinear Wannier functions that are both
real and symmetric (antisymmetric) about $x=0$.

\begin{figure}[!htb]
\includegraphics[width=9.0cm]{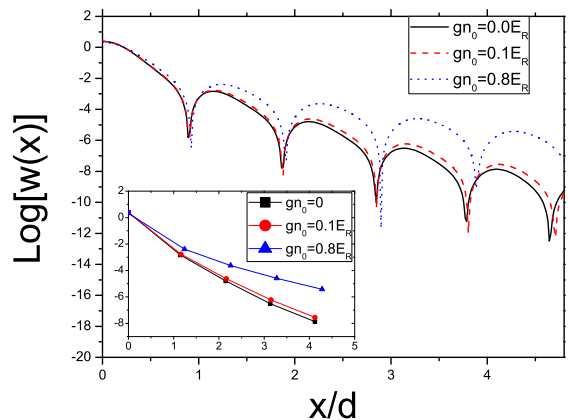}
\caption{(color online)1D nonlinear Wannier functions $w(x)$ for
 different mean-field interactions
$gn_0$. The insets show the decay of the local maximums of the 1D
nonlinear Wannier function.} \label{fig:decay}
\end{figure}

Our  numerical results of the 1D nonlinear Wannier functions are
plotted in a semilog fashion in Fig. \ref{fig:decay}. As clearly
shown, the decay of the the Wannier function remains exponential
despite the nonlinearity. The effect of the nonlinearity (or
interaction) is to make the decay slower and thus the Wannier
function less localized.  The single-particle Wannier function has
been proven to decay exponentially by Kohn\cite{Kohn}. It was
pointed out later \cite{Cloiseaux} that this exponential decay is
related to a well-known mathematical result that connects the
behavior of a function near a branch point to the asymptotic decay
of its Fourier transform \cite{Olver,He}. Suppose that $f_k(x)$ is a
periodic function $f_{k}(x)=f_{k+2\pi/d}(x)$ and has a leading
behavior at the branch point $k_0=\pi/d+ih$ as
$f_k(x)=f_0(x)+\gamma\left[i(k-k_0)\right]^{\beta}$. Its Fourier
transform is an exponential decay function, $F(x)=\int dk
f_{k}(x)e^{-ikx}=2\gamma
\sin\pi(1+\gamma)\Gamma(1+\gamma)x^{-(1+\gamma)}e^{-hx}$. This means
that the nonlinear Wannier functions, shown to decay exponentially
in Fig. \ref{fig:decay},  may also have these analytical properties.
The rigorous proof for this, however, is left for the future
investigation. Also note that due to the power-law prefactor
\cite{He}, the decay in Fig. \ref{fig:decay} is not strictly
exponential as indicated by the slight curving of the envelope of
the peaks.

Nonlinear Wannier functions can be computed similarly for the 2D and
3D optical lattices. However, the computation time can become
enormous in particular for the 3D case. Since our ultimate goal is
to compute the two basic parameters $J$ and $U$ of the BHM  from
Eqs. (\ref{J}) and (\ref{U}), respectively, we can reduce the
computation time significantly by not calculating out the Wannier
function explicitly. For $J$, one can combine Eq.(\ref{J}) and
Eq.(\ref{Wannier}) to express $J$ in terms of the Bloch functions
and then use this expression to compute $J$ efficiently. For $U$,
one can reduce the computing time by utilizing a well-known fact
that $U$ is proportional to the difference between $\mu({\bf k}=0)$
and the system's mean-field ground state energy.

\begin{figure}[tbh]
\includegraphics[width=9.0cm]{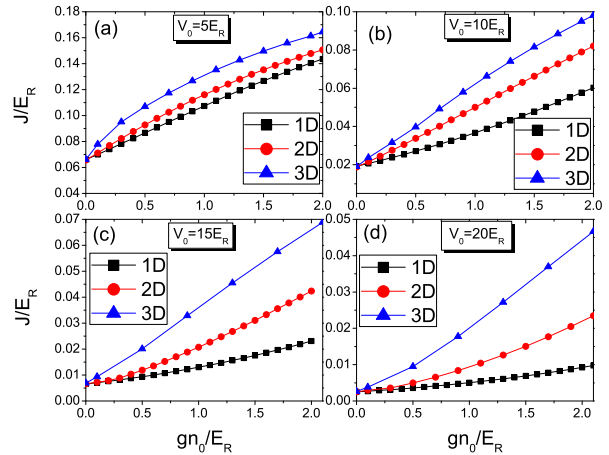}
\caption{(color online)Tunneling rate $J$ via the mean-field interaction $gn_0$
for optical lattices with different strength $V_0$. $1D$, $2D$, and
$3D$ correspond to the dimensionality of the optical lattice. The
Tunneling rate $J$, mean-field interaction $gn_0$ and lattice depth
$V_0$ are all in units of $E_R$.} \label{fig:J}
\end{figure}

\begin{figure}[tbh]
\includegraphics[width=9.0cm]{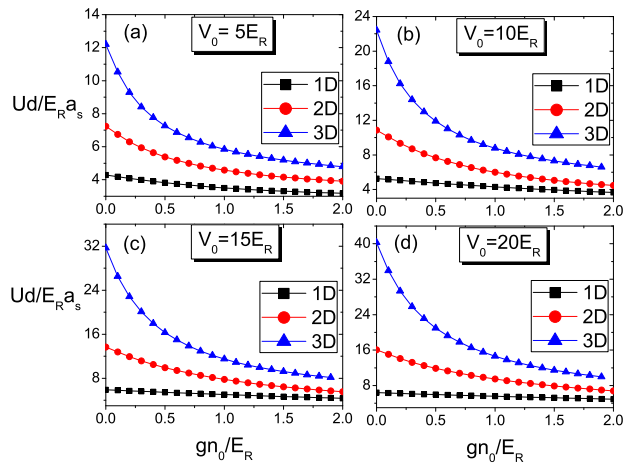}
\caption{(color online) On-site repulsion $U$ via the mean-field interaction $gn_0$
for optical lattices with different strengths  $V_0$. $1D$, $2D$, and
$3D$ correspond to the dimensionality of the optical lattice.
Repulsive on-site interaction $U$ is in units of $E_Ra_s/d$; the
mean-field interaction $gn_0$ and lattice depth $V_0$ are in units
of $E_R$.} \label{fig:U}
\end{figure}

Our numerical results for $J$ and $U$ are shown in
Figs.\ref{fig:J} and \ref{fig:U} for different values of $V_0$ and
$gn_0$. As clearly shown in the figures, the mean-field interactions $gn_0$  have
pronounced effects on both the tunneling rate $J$ and on-site
repulsion $U$. For a fixed lattice strength $V_0$, with the increase of
$gn_0$, $J$ increases while $U$ decreases dramatically.
This is expected as the Wannier function becomes less localized
as $gn_0$ increases.  Moreover, it is apparent from the figures,
in higher dimensions, the interaction effects are much
stronger. This highlights the need to take into account
the interaction effect into $J$ and $U$ since most of the
experiments are carried out with the 3D optical lattices.
Since the optical lattices have
the form $V_{latt}=V_0 E_R\sum_{j=1}^{D}\sin^2\left(q_B
x_j\right)$, the different directions in our system
are decoupled in the linear case $gn_0=0$. The dependence
of the interaction effects on dimensionality shows that
the interaction can strongly couple the motions along
different directions.

Let us take a closer look at a BEC in a 3D optical lattice with
$V_0=10E_R$. At a low filling with $gn_0=0.01E_R$, we have
$J/E_R=0.019$ and $Ud/E_Ra_s=22.4$ (see Figs. \ref{fig:J} (b) and
\ref{fig:U} (b)), which is consistent with the results calculated
with the single-particle Wannier function in Ref. \cite{Xu,Gerbier}.
However, at a higher filling with $gn_0=2.0E_R$, $J$ is
approximately three times larger than the value calculated with the
single-particle Wannier function. Meanwhile, the on-site energy
$Ud/E_Ra_s$ is significantly reduced to $6.6$. This strong
dependence of $J$ and $U$ on the $gn_0$ justifies the necessity of
introducing the nonlinear Wannier functions. In this work, the
parameters $gn_0$ in Figs. \ref{fig:J} and \ref{fig:U} relates to
the filling factor $\left<n_i\right>$, which is often used in the
literature \cite{Morsch,Blochrev2}, as $n_0=\left<n_i\right>/d^3$.

In a recent experiment by Campbell {\it et al.} \cite{Campbell}, the
on-site energy $U$ was measured for different filling factors with
the two-photon Bragg spectroscopy \cite{Stenger}. They found that
$U=22$Hz for the $\left<n_i\right>=5$ shell at $V_0=35E_R$, a
decrease of 27\% from $U=30$Hz for $\left<n_i\right>=1\sim 2$. Our
numerical results are $U=28.4$Hz for $\left<n_i\right>=5$ and
$U=33.1\sim 31.8$Hz for $\left<n_i\right>=1\sim 2$, in good
agreement with the experiment. This shows that even though the
nonlinear Wannier function is a mean-field concept, it somehow still
captures much of the interaction effect  in the Mott insulator
regime. A possible method for the experimental study on the
tunneling rate $J$  is to study the interference pattern produced by
an expanding atomic cloud \cite{Gerbier}.

We emphasize here that our calculation of the two basic parameters
of the BHM has been done with  the mean-field theory. Further
improvement of the theoretical framework is also needed to include
the effects of quantum fluctuations \cite{Petrov}.

To conclude, we have demonstrated a way to
construct a set of orthonormal Wannier functions by properly
incorporating interaction effect for a BEC in an optical lattice.
Although these nonlinear Wannier functions are less localized
than the single-particle Wannier functions due to the repulsive
interaction, they retain many
of the analytical properties of the single-particle Wannier functions.
For example, our numerical results show that they decay
exponentially. We have used these Wannier functions
to compute the tunneling rate $J$ and the on-site interaction
$U$ in the Bose-Hubbard model. The computed $U$ are found
in good agreement with experimental results.

We thank Ying Hu for helpful discussions. This work was supported in
part by grants from the Hongkong Research Council (RGC) and the Hong
Kong Baptist University Faculty Research Grant (FRG). Z.X.L. was
supported by the IMR SYNL-T.S. K$\hat{e}$ Research Fellowship. B.W.
was supported by the NSF of China (10825417) and the 973
project of China (2005CB724500,2006CB921400).

\end{document}